\documentclass[twocolumn]{webofc}

\usepackage{bm}
\usepackage[colorlinks,citecolor=blue]{hyperref}
\begin{document}
\title{Skyrme-RPA study of charged-current neutrino opacity in hot and dense supernova matter}

\author{\firstname{Alan} \lastname{A. Dzhioev}\inst{1}\fnsep\thanks{\email{dzhioev@theor.jinr.ru}} \and
           \firstname{Gabriel} \lastname{Mart\'{\i}nez-Pinedo}\inst{2,3}\fnsep\thanks{\email{g.martinez@gsi.de}}
        }

\institute{
 Bogoliubov Laboratory of Theoretical Physics, JINR, 141980, Dubna, Russia
\and
 GSI Helmholtzzentrum f\"ur Schwerionenforschung, Planckstra{\ss}e~1, 64291 Darmstadt, Germany
\and
Institut f{\"u}r Kernphysik
  (Theoriezentrum), Technische Universit{\"a}t Darmstadt,
  Schlossgartenstra{\ss}e 2, 64289 Darmstadt, Germany}

\abstract{%
  Neutrino emission and their transport of energy to the supernova
  shock region are sensitive to the physics of hot and dense nuclear
  matter, which is a complex problem due to the strong correlations
  induced by nuclear forces.  We derive charged-current opacities for
  electron neutrinos and antineutrinos in supernova matter using a
  self-consistent approach based on the Skyrme effective
  interaction. We include a mean field energy shift due to nuclear
  interaction and corrections due to RPA correlations.  A complete
  treatment of the full Skyrme interaction, including a spin-orbit
  term, is given.  To test the effect of RPA corrections, neutrino and
  antineutrino opacities are computed using different Skyrme
  parametrizations consistent with a number of infinite matter
  constraints.}
\maketitle
\section{Introduction}
\label{intro}

Neutrino and antineutrino absorption reactions in nuclear matter play
a crucial role in core-collapse
supernovae~\cite{Reddy.Prakash.Lattimer:1998,Burrows.Reddy.Thompson:2006}.
In delayed neutrino-heating mechanism, the energy transport by
electron neutrinos and antineutrinos from a hot protoneutron star
(PNS) is responsible for reviving the stalled shock-wave and
generating an explosion~\cite{Bethe1985,Janka2012}. The
charged-current reactions $\nu_e + n \to p + e$ and
$\bar\nu_e + p \to n + e^+$ provide a dominant mechanism for heating
the matter behind the shock. The efficiency of neutrino-heating
mechanism and the success of the explosion depend critically on the
luminosity and energy spectra for $\nu_e$ and $\bar\nu_e$. In turn,
these characteristics are determined by the neutrino transport inside
the protoneutron star, and thus depend on the neutrino opacities in
the dense ($\rho\simeq10^{12}-10^{14}~\mathrm{g/cm^3}$) and hot
($T\simeq 5-10~\mathrm{MeV}$) nuclear matter. The main opacity sources
for $\nu_e$ and $\bar\nu_e$ are the same absorption reactions on
nucleons, which are responsible for the neutrino-heating mechanism.

For dense neutron-rich conditions relevant in the protoneutron star,
earlier calculations have found that many-body correlations due to
strong interactions between nucleons can modify the neutrino opacities
significantly. In particular, it was
found~\cite{MartinezPinedo_2012,Roberts_PRC2012} that the difference
in the neutron and proton mean-field potential energies enhances
(suppresses) the (anti)neutrino opacity considerably with respect to
that in the noninteracting gas of free nucleons. Meanwhile,
in~\cite{Roberts_PRC2012} it was shown that random-phase approximation
corrections, accounting for long-range $ph$ correlations, act in the
opposite direction, but their effect is less significant as compared
to that of mean-field potential difference.

It was emphasized in many works (see, for example, Refs.~\cite{Sawyer_PRD11,Sawyer_PRC40,Reddy_PRC59}) that neutrino opacities need to be consistent with the model
of the strong interaction used in the equation of state for protoneutron star matter.  Specifically, a particular choice of the particle-hole interaction should be
consistent with the EOS employed to compute the composition of charge-neutral $\beta$-equilibrated matter~\cite{Reddy_PRC59}. In this paper we use the EOS based on the Skyrme
effective interaction to compute $\nu_e$ and $\bar\nu_e$ charged-current opacities in the outer region of PNS called the neutrino-sphere.  We employ self-consistent RPA calculations,
i.e., both the Hartree-Fock mean-field   and the residual $ph$ interaction are derived from the same Skyrme potential. In this work we will consider five Skyrme parameterizations:
KDE0v1, LNS, NRAPR, SKRA, SQMC700. It was shown in Ref.~\cite{Dutra_PRC85} that these Skyrme forces satisfy a series of  experimentally extracted constraints for infinite nuclear matter properties.

The paper is organized as follows. In Sec.~\ref{formalism} we briefly present the Skyrme-RPA formalism to compute the Fermi and Gamow-Teller strength functions which
determine the absorption cross section for $\nu_e$ and $\bar\nu_e$.  Our results for the Fermi and Gamow-Teller strength functions as well as for (anti)neutrino opacities are presented in Sec.~\ref{results}. Finally, we conclude in Sec.~\ref{conslusions}.

\section{Formalism}\label{formalism}

The charged-current opacity for $\nu_e$ and $\bar\nu_e$  can be calculated by numerical integration of the absorption differential cross section per unit volume.
In the nonrelativistic limit for nucleons, the double-differential cross-section for a  (anti)neutrino with initial energy $E_\nu$  is given by~\cite{Roberts_PRC2012}
\begin{align}\label{CrSect}
 & \frac{1}{V}\frac{d^2\sigma}{d\cos\theta\,dE_e}=\frac{G^2_F \cos^2\theta_C}{2\pi}p_e E_e (1-f_e(E_e))
  \\ \notag
 & \times\bigl[(1+\cos\theta)S^{(\pm)}_0(\omega, q) + g^2_A (3 - \cos\theta)S^{(\pm)}_1(\omega, q) \bigr].
\end{align}
The energy transfer to the nuclear medium is $\omega = E_\nu - E_e$,
while the momentum transfer is $q=|\bm{q}|=|\bm{p}_\nu - \bm{p}_e|$.
$S^{(\pm)}_0$ and
$S^{(\pm)}_1$ are the strength functions (also called dynamical
structure factors) associated with the Fermi and Gamow-Teller
operators, $\bm{\tau}^{(\pm)}$ and $\bm{\sigma\tau}^{(\pm)}$,
respectively. Note, that in Eq.~\eqref{CrSect} the plus sign refers to
a neutrino absorption, while the minus sign refers to an antineutrino
absorption. Thus, for (anti)neutrino absorption the cross section
includes Fermi and Gamow-Teller strength functions for $n\to p$
($p\to n$) transitions. Finally, $1-f_e(E_e)$ is the Pauli blocking
factor for the outgoing lepton. In the protoneutron star matter
positrons are nondegenerate and their spectrum approaches the
Boltzmann distribution. Therefore, we can neglect the final-state
blocking for antineutrino absorption.

The strength functions $S^{(\pm)}_{0,1}$ embody all spatial and
temporal correlations between nucleons arising from strong
interaction.  At $T\ne 0$ they include summation of transition
probabilities from thermally excited states $|i\rangle$ to final
states $|f\rangle$ such that $\omega = E_f-E_i$:
\begin{align}
  S^{(\pm)}_\alpha(q,\omega) = \sum_{i,f} P_i|\langle f |\bm{Q}^{(\pm)}_\alpha| i \rangle|^2\delta(E_f-E_i - \omega),
\end{align}
with $P_i=Z^{-1}\mathrm{e}^{-E_i/T}$ being the probability of finding
the system in the state $|i\rangle$, and $Z=\sum_ie^{-E_i/T}$
is the partition function.  The one-body transition operators
$\bm{Q}^{(\pm)}_\alpha$ are defined as $\bm{Q}^{(\pm)}_0  = \sum_j e^{i\bm{q}\bm{r}_j}\bm{\tau}^{(\pm)}_j$, $ \bm{Q}^{(\pm)}_1  = \sum_j e^{i\bm{q}\bm{r}_j}\bm{\sigma}_j\bm{\tau}^{(\pm)}_j,$
where the index $j$ stands for the nucleon. The strength functions obey the detailed balance theorem~\cite{Hernandez_NPA658}
\begin{equation}\label{DB}
  S^{(\pm)}_\alpha(q,\omega) =  e^{(\omega \pm \Delta \mu_{np})/T} S^{(\mp)}_\alpha (q,-\omega),
\end{equation}
where $\Delta \mu_{np} = \mu_n - \mu_p$ is the difference between the neutron and proton chemical potentials,
as well as they satisfy the non-energy weighted sum rule~\cite{Hernandez_NPA658}
\begin{equation}\label{sum_rule}
 \int^{+\infty}_{-\infty} d\omega\bigl[ S^{(+)}_\alpha(q,\omega) -  S^{(-)}_\alpha(q,\omega)\bigr] = \rho_n - \rho_p,
\end{equation}
which depends only on the difference  between neutron and proton densities.

We obtain the strength functions from the imaginary part of the response function $\Pi^{(\alpha)}(q,\omega) $ according to the   fluctuation-dissipation theorem
\begin{equation}
   S^{(\pm)}_\alpha(q,\omega) = -\frac{1}{\pi}\frac{\mathfrak{Im}
    \Pi^{(\pm)}_\alpha(q,\omega)}{1-\mathrm{e}^{-(\omega \pm \Delta \mu_{np})}}.
\end{equation}
In its turn $\Pi^{(\pm)}_\alpha$ is related to the particle-hole propagator (or the retarded $ph$ Green's function)
\begin{equation}
  \Pi^{(\pm)}_\alpha(q,\omega) = 2 \int\frac{d^3\bm{k}}{(2\pi)^3} G^{(\pm)}_\alpha(k,q,\omega).
\end{equation}
Here and below, for the momentum averages  we adopt the notation
\begin{equation}
  \int\frac{d^3\bm{k}}{(2\pi)^3} f(\bm{k})G^{(\pm)}_\alpha(k,q,\omega) \equiv\langle f G^{(\pm)}_\alpha\rangle.
\end{equation}
With this notation $ \Pi^{(\pm)}_\alpha(q,\omega) = 2\langle G^{(\pm)}_\alpha\rangle$.

Neglecting long-range $ph$ correlations  $G^{(\pm)}_\alpha = G^{(\pm)}_{HF} $, and the Hartree-Fock particle-hole propagator is given by
\begin{equation}
G^{(\pm)}_{HF}(k, q, \omega) = \frac{ f_{n(p)}(\bm{k}) -f_{p(n)}(\bm{k} + \bm{q})}{\omega + E_{n(p)}(\bm{k}) - E_{p(n)}(\bm{k} + \bm{q}) + i\eta }.
\end{equation}
Here $f_\tau(\bm{k})=\bigl[1+\mathrm{e}^{(E_\tau(\bm{k})-\mu_\tau)/T}\bigr]^{-1}$ is the Fermi-Dirac distribution for nucleons and  $E_\tau(\bm{k})$ is the single particle energy given
by
\begin{equation}\label{disp}
  E_\tau(\bm{k}) = \frac{\bm{k}^2}{2m_\tau} + U_\tau,
\end{equation}
where $m_\tau$ is the effective mass and $U_\tau$ is the Hartree-Fock potential obtained with the Skyrme interaction.
Using the detailed balance~\eqref{DB} we can write the Hartree-Fock strength function for $n\to p$ ($p\to n$) transitions as
\begin{align}\label{S_HF}
  S^{(\pm)}_{HF} (q,\omega)= 2\int \frac{ d^3\bm{k}}{(2\pi)^3}\,\delta(E_{p(n)} - E_{n(p)}-\omega)
  \notag\\
  \times f_{n(p)}( \bm{k}) (1-f_{p(n)}(\bm{k} + \bm{q})).
\end{align}
Assuming $U_\tau=0$ and considering the bare nucleon masses we get the strength function for a noninteracting Fermi gas.

To go beyond the HF approximation we take into the residual proton-neutron interaction by resumming a class  of so-called ring diagrams and obtain the well known RPA~\cite{Fetter}.
The RPA correlated  $ph$ propagator is the solution of the Bethe-Salpeter integral equation
\begin{align}\label{BS_eq}
  G^{(\pm)}_{\alpha}(k_1, q,  \omega) =  G^{(\pm)}_{HF} (k_1, q,  \omega) + G^{(\pm)}_{HF} (k_1, q,  \omega)
  \notag\\
 \times\sum_{\alpha'}\int\frac{ d^3\bm{k_2}}{(2\pi)^3} V^{(\alpha,\alpha')}_{ph}(\bm{q},\bm{k}_1,\bm{k}_2) G^{(\pm)}_{\alpha'}(k_2, q,  \omega),
\end{align}
where $V^{(\alpha,\alpha')}_{ph}$ is the residual interaction matrix element. For the Skyrme effective interaction containing a zero-range spin-orbit term this proton-neutron matrix element
takes the form~\cite{Davesne_PRC89}
\begin{multline}
V_{ph}^{(SM;S'M')}(\bm{q},\bm{k}_1,\bm{k}_2) \equiv
\\
  \langle \bm{q}+\bm{k}_1,\bm{k}_1^{-1};SM|V_{ph}|\bm{q}+\bm{k}_2,\bm{k}_2^{-1};S'M'\rangle =
\\
 \delta_{SS'}\delta_{MM'}\bigl(W^{(S)}_1 + W^{(S)}_2\bm{k}^2_{12} \bigl)
 \notag \\
+ q W^{(1)}_{SO}\bigl(\delta_{S'0}\delta_{S1}M(k_{12})^{(1)}_{-M} + \delta_{S'1}\delta_{S0}M'(k_{12})^{(1)}_{M'}\bigr),
\end{multline}
where $\bm{k}_{12} = \bm{k}_1 - \bm{k}_2$ is the relative hole
momentum, while the rank-1 tensor $(k_{12})^{(1)}_{M}$ is defined
in~\cite{Davesne_PRC89}.  The coefficients $W^{(S)}_{1,2}$ are the
combinations of the Skyrme parameters $(x_i, t_i)$, the transferred
momentum $q$ and the total density $\rho=\rho_n+\rho_p$. Their
detailed expressions are given in~\cite{Hernandez_NPA658}. It is
important to note that the spin-orbit term introduces a coupling
between $S=0$ and $S=1$ spin channels and removes the degeneracy on
the spin projection $M$.

Following the method suggested by Margueron~\textit{et
  al}~\cite{Margueron_PRC74}, a closed algebraic system for
$\langle G^{(\pm)}_\alpha\rangle$ is obtained by multiplying the
Bethe-Salpeter equation~\eqref{BS_eq} with the functions $k^2_1$,
$k_1 Y_{1\mu}$ and integrating over the momentum. Finally, the system
of algebraic equations can be written in a compact form for each
channel ($\alpha=S,M$) as
\begin{align}\label{BS1}
  &(1-\widetilde W_1^{(\alpha)}\beta_0^{(\pm)} - W_2^{(\alpha)} q^2\beta_2^{(\pm)})\langle G^{(\pm)}_\alpha\rangle
  \notag\\
  &- W_2^{(\alpha)}\beta_0^{(\pm)}\langle k^2 G^{(\pm)}_\alpha\rangle+ 2 W_2^{(\alpha)}q\beta_1^{(\pm)}
  \notag\\
  &\times\sqrt{\frac{4\pi}{3}}\langle k Y_{10} G^{(\pm)}_\alpha\rangle = \beta_0^{(\pm)},
  \end{align}
  \begin{align}\label{BS2}
    -(&\widetilde W_1^{(\alpha)}q^2\beta_2^{(\pm)} + W_2^{(\alpha)} q^4\beta_5^{(\pm)})\langle G^{(\pm)}_\alpha\rangle
    \notag\\
    &+(1-W_2^{(\alpha)}q^2\beta_2^{(\pm)})\langle k^2 G^{(\pm)}_\alpha\rangle + 2 W_2^{(\alpha)}q^3\beta_4^{(\pm)}
    \notag\\
    &\times\sqrt{\frac{4\pi}{3}}\langle k Y_{10} G^{(\pm)}_\alpha\rangle = q^2\beta_2^{(\pm)},
  \end{align}
  \begin{align}\label{BS3}
    -(&\widetilde W_1^{(\alpha)}q\beta_1^{(\pm)} + W_2^{(\alpha)} q^3\beta_4^{(\pm)})\langle G^{(\pm)}_\alpha\rangle
    \notag\\
    &- W_2^{(\alpha)}q\beta_1^{(\pm)}\langle k^2 G^{(\pm)}_\alpha\rangle + (1+2 W_2^{(\alpha)}q^2\beta_3^{(\pm)})
    \notag\\
    &\times\sqrt{\frac{4\pi}{3}}\langle k Y_{10} G^{(\pm)}_\alpha\rangle = q\beta_1^{(\pm)}.
\end{align}
The parameter $\widetilde W_1^{(\alpha)}$  describes the coupling between spin channels induced by the
spin-orbit interaction
\begin{equation}
\widetilde W_1^{(\alpha)} =  W_1^{(\alpha)} + C^{(\alpha)}\frac{W^2_{SO}q^4(\beta_2^{(\pm)}-\beta_3^{(\pm)})}{1+W_2^{(\alpha')}q^2(\beta_2^{(\pm)}-\beta_3^{(\pm)})}.
\end{equation}
Here $\alpha'$ is defined with respect to $\alpha$ as $S'=1-S$, while  $C^{(\alpha)}=1$ if $S=0$ and $C^{(\alpha)}=\frac12 M^2 $ if $S=1$.
The six functions  $\beta^{(\pm)}_l(q,\omega),~(l=0,5)$ include the momentum averages of the HF propagator $G^{(\pm)}_{HF}$  and their
 explicit expressions are given ~\cite{Hernandez_NPA658}.
 If we replace $\widetilde W_1^{(\alpha)}$  in Eqs.~(\ref{BS1}-\ref{BS3}) by $W_1^{(\alpha)}$ we obtain the results of Ref.~\cite{Hernandez_NPA658}, related to the central part of the interaction,
as it should be. We also would like to note that for ambient conditions (density and temperature) considered in the present work, the effect of the spin-orbit  interaction is negligible and
for Gamow-Teller strength function we have $S^{(\pm)}_{S=1,M=0}\approx S^{(\pm)}_{S=1,M=1}\equiv S^{(\pm)}_{S=1}$.

\section{Results}\label{Results}\label{results}

\begin{figure}[t]
\centering
\includegraphics[width=\columnwidth]{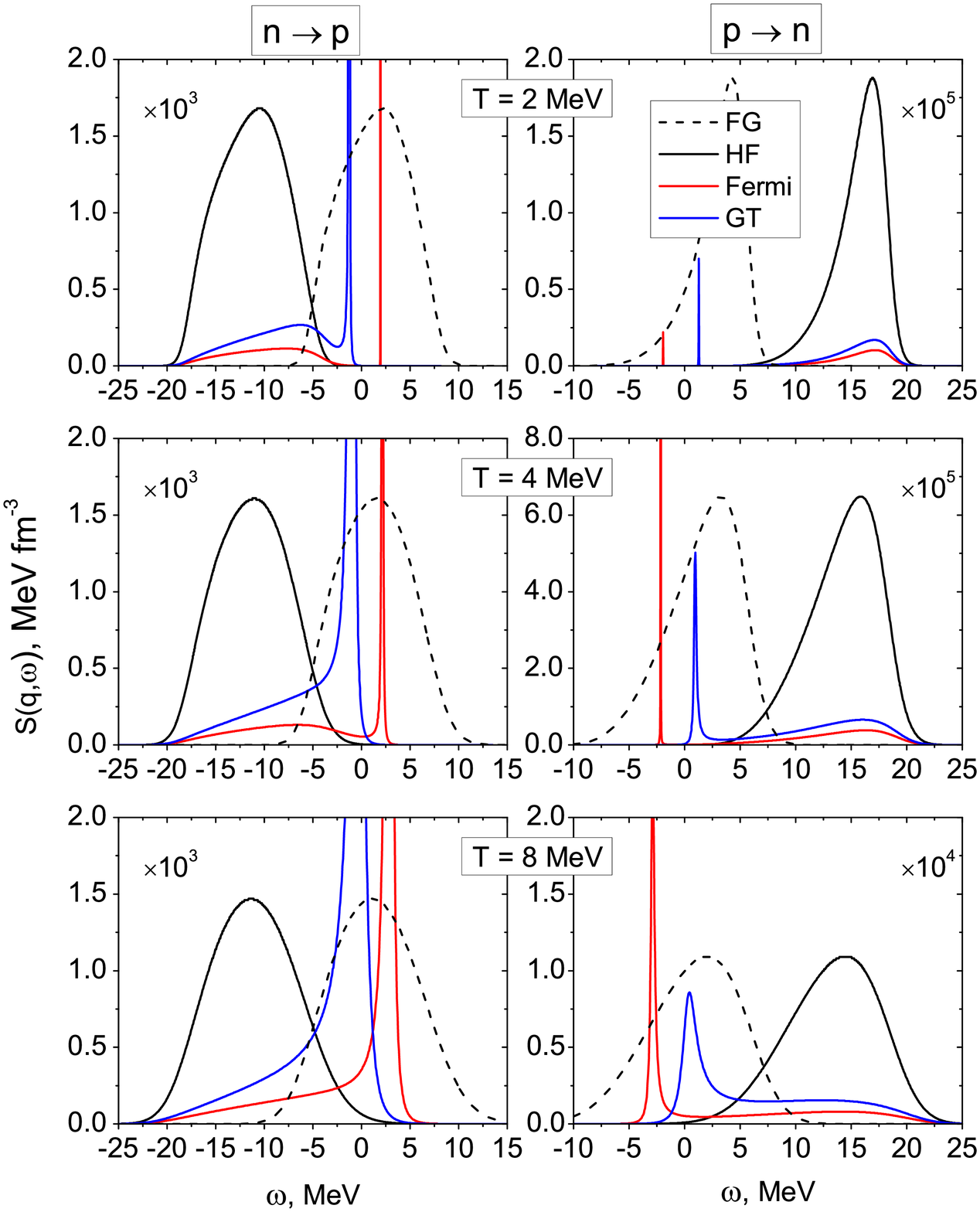}
\caption{The finite temperature ($T=2,\,4,8$~MeV) HF and RPA strength functions at $q=0.2\,\mathrm{fm}^{-3}$  as  functions of energy $\omega$ for $n\to p$ (left panels) and $p\to n$ (right panels) transitions. The dashed curves give the strength function for the noninteracting Fermi gas (FG).  Note, that the strength functions are scaled by the factor shown inside each panel.}
\label{fig-1}       
\end{figure}

To illustrate the general features of the strength functions we have
chosen the LNS Skyrme parametrization. The temperature evolution of
Fermi and GT strength functions is shown in Fig.~\ref{fig-1} for both
$n\to p$ and $p\to n$ transition at a small momentum transfer
$q=0.2$~fm$^{-1}$. We have performed calculations at density
$\rho=0.02$~fm$^{-3}$ for a proton fraction
$Y_p=\rho_p/\rho = 0.1$.  As a reference, the HF and noninteracting
Fermi gas (FG) strength functions -- which are independent of the spin
channel - are also shown.

Comparing the noninteracting $n\to p$ FG strength functions with the HF ones we find that they have the same shape, but the interaction shifts the HF strength function to larger negative  energy transfer. It means
that the energy of the nuclear system is decreased for $n\to p$ transitions. In contrast,  the interaction pushes the HF $p\to n$ strength function to higher energy transfer,
thereby increasing the  energy of the nuclear system. This is understood as follows. The dispersion relation for the nucleon is given by Eq.~\eqref{disp} and the energy shift $\Delta U_{np}=U_n-U_p$ is closely related to the symmetry energy, which describes how the energy of nuclear matter changes as one moves away
from the equal number of protons and neutrons. For the LNS Skyrme interaction the potential difference at considered neutron-rich conditions
is $\Delta U_{np}\approx 12$~MeV. Thus the proton is more strongly bound than the neutron and the difference in potentials serves as a threshold for the $p\to n$ reactions.

Allowing for repulsive $ph$ interaction at low temperatures
significantly suppresses the continuum part of the HF strength
functions and gives rise to well-defined Fermi and Gamow-Teller
collective modes either at positive or negative energy above the
particle-hole band. Moreover, due to the detailed balance~\eqref{DB},
each collective mode peak in the $n\to p$ strength function has its
image in the $p \to n$ strength function at $-\omega$. One important
point is that these $p\to n$ collective modes appear only due to
thermal effects and they are located below the particle-hole band.
The centroid of collective excitations is almost stable against
thermal effects, while their width grows with increasing temperature
due to Landau damping.  At high temperatures collective modes become
absorbed into thermally extended particle-hole band.

\begin{figure}[t]
\centering
\includegraphics[width=\columnwidth]{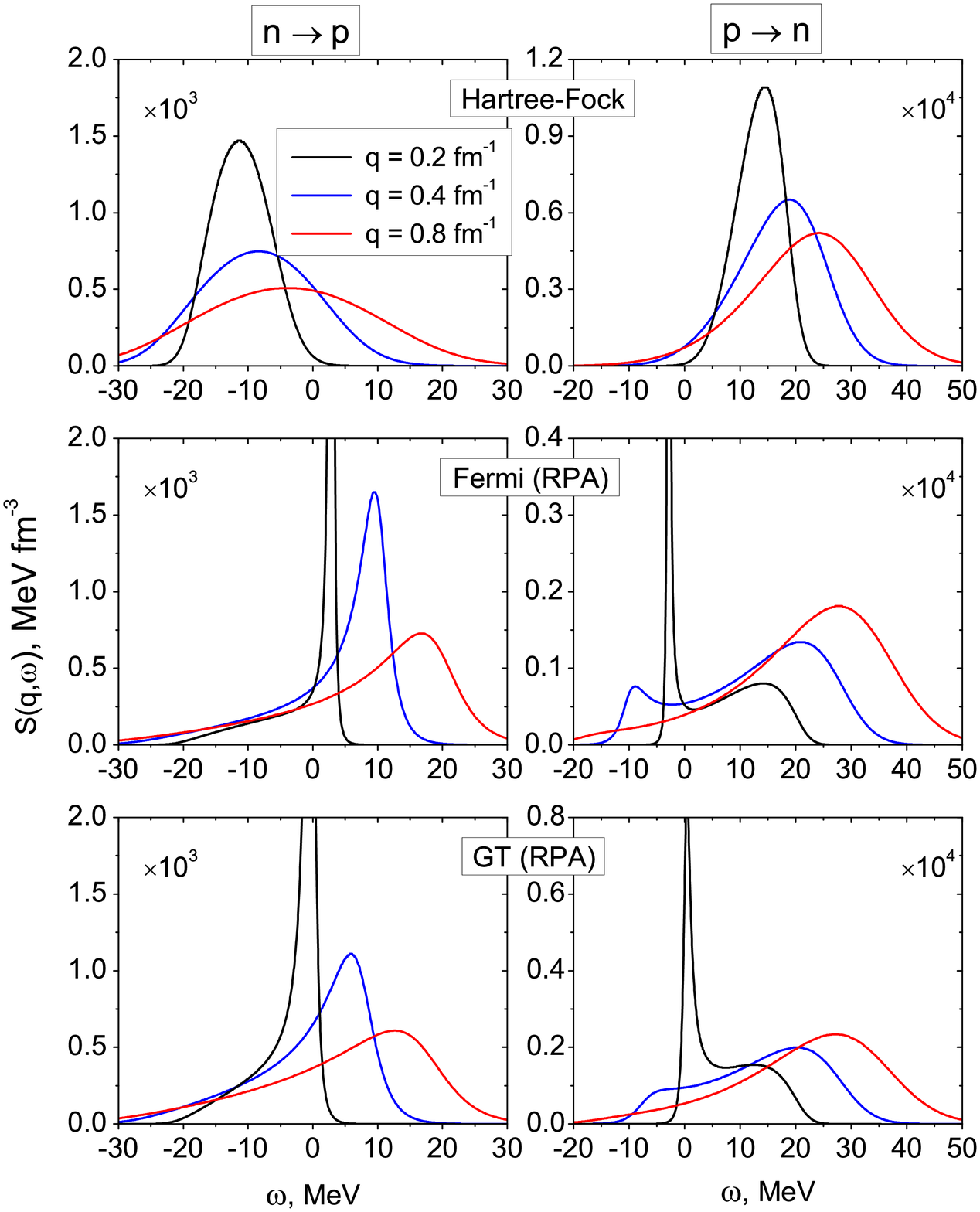}
\caption{Finite temperature ($T=8$~MeV) strength functions calculated with the LNS Skyrme potential at three values of momentum transfer. The density and the proton fraction are the same as in Fig.~\ref{fig-1} }
\label{fig-2}       
\end{figure}

Figure~\ref{fig-2} shows a typical modification of the Fermi and
Gamow-Teller strength functions, $S^{(\pm)}_0$ and $S^{(\pm)}_{1}$,
with increasing momentum transfer. On the same figure the HF strength
functions are shown for reference. As we have mentioned before and as
it follows from Eq.~\eqref{S_HF}, the evolution of the HF strength
functions for $n\to p$ ($p\to n$) transitions follows that of the
noninteracting Fermi gas, but shifted by $\Delta U_{np}\approx 12$~MeV
to lower (higher) energies. In particular, we can observe the
enlarging the particle-hole band with increasing momentum transfer $q$
following the overall reduction of the strength. The sum
rule~\eqref{sum_rule} requires that for neutron-rich conditions the
momentum increase has a smaller effect on the total $n\to p$ strength
than on that for the $p\to n$ transitions.

The middle and lower panels of Fig.~\ref{fig-2} show that  the effects of increasing momentum  are remarkably similar for the  Fermi and Gamow-Teller strength functions.
At low momentum transfer the $n\to p $ ($p\to n$) RPA strength functions in both the $S=0$ and $S=1$ channels have relatively narrow peaks  above (below) the  maximum
of the HF strength.  When $q$ increases, the peaks become broader and finally the RPA strength evolves to a structureless shape. Furthermore, although at large $q$ the  $ph$ correlations
have a relatively smaller effect than at small $q$, the $n\to p $ RPA strength functions  remain shifted to larger energies with respect to the HF strength. In contrast, because of diminishing contribution from thermally excited collective modes, the maximums  of the $p\to n $  Fermi and Gamow-Teller strength functions move from low energies to higher ones and finally they appear above the HF maximum.

\begin{figure*}
\centering
\includegraphics[width=\textwidth,clip]{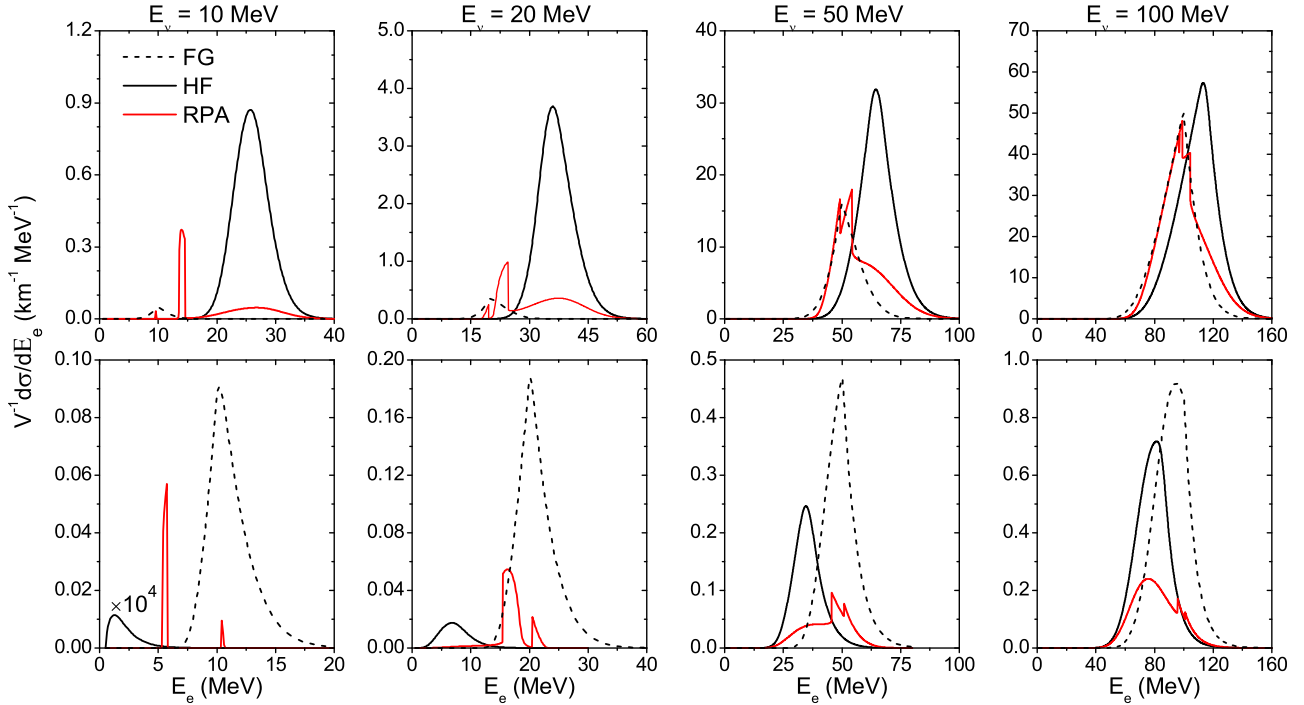}
\caption{Differential cross sections for neutrino (upper panels) and antineutrino (lower panels) absorption, evaluated at beta equilibrium conditions
$\rho=0.02\,\mathrm{fm}^{-3}$, $T=8\,\mathrm{MeV}$, and $Y_p = 0.029$. The Skyrme force LNS is used in the HF and RPA calculations.
In each panel the dashed line shows the noninteracting Fermi gas cross sections. For $E_{\bar\nu_e}=10$ antineutrinos the HF cross section is scaled by the factor $10^4$. }
\label{fig-3}       
\end{figure*}

The differential absorption cross sections for neutrinos and antineutrinos as a function of the outgoing lepton energy are shown in Fig.~\ref{fig-3}.
We consider incoming neutrino energies $E_{\nu,\bar\nu}=10,~20,~50$ and $100$~MeV and the conditions of the medium are  $\rho=0.02\,\mathrm{fm}^{-3}$,
$T=8\,\mathrm{MeV}$. The proton fraction and the electron chemical potential are $Y_p = 0.029$ and $\mu_e=47.05\,\mathrm{MeV}$, respectively, which correspond to beta-equilibrium (i.e., $\mu_e=\mu_n-\mu_p$) for the given temperature, density and the assumed
nuclear Skyrme interaction LNS. The mean-field potential difference is $\Delta U_{np}=14.66\,\mathrm{MeV}$.

It is seen from the figure, the peak of the differential cross section
is shifted by about $\Delta U_{np}$ to higher (lower) energies for
(anti)neutrino absorption with respect to the noninteracting Fermi gas
model and, hence, the mean-field potential contribution increases
(reduces) the energy of the emitted electron (positrons). This
observation is in line with the elastic
approximation~\cite{Reddy.Prakash.Lattimer:1998} where the energy of
the outgoing electron (positron) is $E_e=E_\nu +\Delta U_{np}$
($E_{e^+}=E_{\bar\nu} -\Delta U_{np}$).  For low energy neutrinos,
$E_\nu<\mu_e$, the additional energy $\Delta U_{np}$ is not enough to
overcome the Pauli blocking for the outgoing electron. However, this
energy is sufficient to put the outgoing electron in a less blocked
portion of phase space, thereby leading to an exponential enhancement
of the cross section. For high-energy neutrinos, $E_\nu>\mu_e$, the
cross section is approximately proportional to the phase space factor
$p_eE_e$ and the effects of the mean field contribution diminish with
increasing $E_\nu$.

For the antineutrino HF cross section the important difference  from the neutrino case comes from the fact that the reaction threshold is increased by $\Delta U_{np}$. The increased reaction threshold leads to a significant suppression of the HF cross section for low-energy
antineutrinos. For $E_{\bar\nu} < \Delta U_{np} $ antineutrinos the HF cross section differs from zero  due to thermal effects which smear the proton and neutron Fermi surfaces.
As was stated above, we can neglect the Pauli blocking for the outgoing positron and for  $E_{\bar\nu} > \Delta U_{np}$ antineutrinos the
HF cross section is suppressed solely by the reduced available phase space.

Let us now consider the role of RPA correlations on the differential cross sections. It is seen from the upper panels of Fig.~\ref{fig-3} that RPA correlations suppress the  neutrino absorption cross section. The degree of suppression depends on neutrino energy and it is most significant for low-energy  neutrinos when the transferred momentum is also low and the RPA strength functions are dominated by the resonance contributions (see~Fig.~\ref{fig-2}). The Gamow-Teller and Fermi resonances are clearly visible in the differential cross section for low-energy neutrinos, however their contribution is strongly suppressed  by the phase-space blocking for outgoing electrons. With increasing  neutrino energy, the contribution of collective states to the $\nu_e$ cross section decreases and the RPA corrections become smaller. Concerning $\bar\nu_e$ absorption, for low-energy antineutrinos the HF cross section is almost zero due to reaction threshold.   However, RPA correlations and thermal effects give rise to collective peaks which decrease the reaction threshold and enhance the cross section significantly.  The contribution of collective states diminishes with increasing antineutrino energy and the RPA corrections tend to reduce the cross section as in the $\nu_e$ case.

\begin{figure}[t]
\centering
\includegraphics[width=\columnwidth]{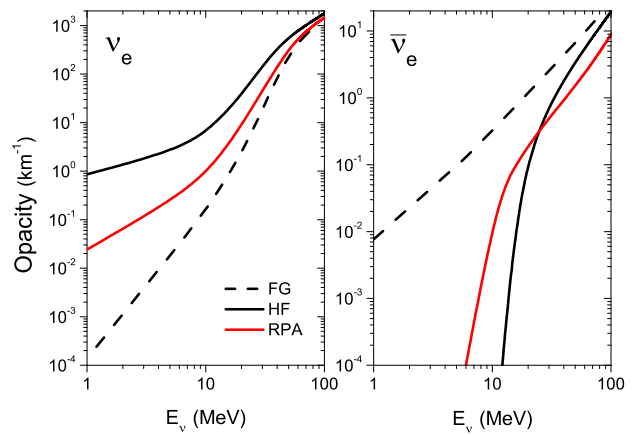}
\caption{Neutrino (left panel) and antineutrino (right panel)
  opacities, evaluated at beta-equilibrium conditions
  $\rho = 0.02\,\mathrm{fm}^{-3}$, $T=8\,\mathrm{MeV}$, and
  $Y_p=0.029$ with the Skyrme force LNS.}
\label{fig-4}       
\end{figure}

Figure~\ref{fig-4} shows neutrino and antineutrino opacities evaluated
at the same conditions considered in Fig.~\ref{fig-3}. The results
shown in Fig.~\ref{fig-4} follow the trends expected from the results
for the differential cross sections. Namely, due to presence of the
mean-field potentials the opacity for low-energy neutrinos increases
with respect to the noninteracting case, but at larger neutrino
energies the mean-field contribution becomes less important and the
opacities computed with and without mean-field potentials approach
each other asymptotically. In contrast, the antineutrino HF opacity is
reduced relative to the Fermi gas result and the reaction threshold
manifests itself at $\Delta U_{np}$. For incoming (anti)neutrino
energy $E_\nu=24$~MeV which is the mean thermal energy $E_\nu\sim 3T$
at temperature $T=8$~MeV the HF neutrino opacity is enhanced by about
an order of magnitude compared to the noninteracting Fermi gas, while
the antineutrino opacity is suppressed by almost a factor of
$5$. While mean-field effect tends to increase the neutrino opacity,
RPA correlations decrease it. For $E_\nu=24$~MeV neutrinos, the RPA
opacity is suppressed by almost a factor of 4 relative to the HF
one. For antineutrino absorption, the inclusion of RPA correlations
decreases the reaction threshold and increases opacity for low-energy
neutrinos. However, at higher neutrino energies RPA correlations
reduces the opacity.

\begin{figure}[t!]
\centering
\includegraphics[width=\columnwidth]{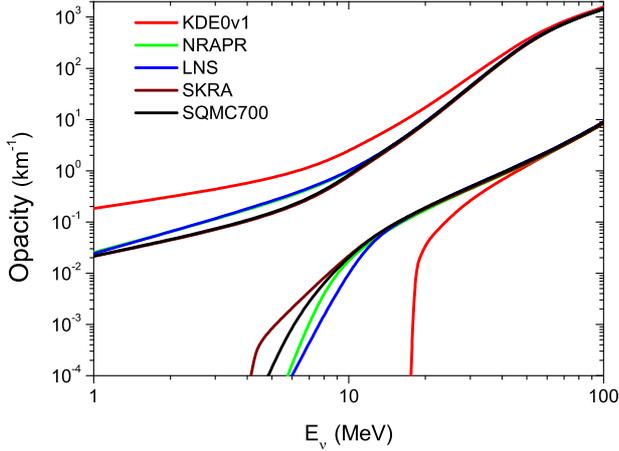}
\caption{Opacity for neutrino and antineutrino, evaluated at conditions $\rho = 0.02\,\mathrm{fm}^{-3}$, $T=8\,\mathrm{MeV}$, and $Y_p=0.029$ with the different Skyrme parameterizations.}
\label{fig-5}       
\end{figure}

We now compare the opacities evaluated with different Skyrme forces mentioned in the Introduction. Results for the parametrizations KDE0v1, LNS, NRAPR, SKRPA, and SQMC700 are shown in Fig.~~\ref{fig-5}.
The RPA opacities  do not differ significantly between the last four Skyrme parametrizations. In contrast,
the HF calculations  based on the KDE0v1 force predict
a mean-field potential difference $\Delta U_{np}=20.7\,\mathrm{MeV}$ which is substantially larger than that for other four parametrizations. As a result, low-energy  neutrino (antineutrino) opacities evaluated with  the KDE0v1 force turn out to be well above (below) those expected from other Skyrme forces.

\section{Conclusions}\label{conslusions}

In this paper, we have studied the combined effects of the mean-field
potential difference and RPA correlations on the $\nu_e$ and
$\bar\nu_e$ opacity in the matter of protoneutron stars.  By solving
the Bethe-Salpeter equation with the residual particle-hole
interaction derived from a Skyrme potential, we estimated the Fermi
and Gamow-Teller strength functions which are directly related to the
(anti)neutrino absorption cross section. Strength functions are
analyzed for varying conditions of momentum transfer and temperature
for a representative Skyrme potential LNS. It is shown that both
mean-field effects and RPA correlations play a significant role in the
strength function redistribution relative to the noninteracting Fermi
gas. In agreement with the previous
studies~\cite{MartinezPinedo_2012,Roberts_PRC2012}, the mean-field
corrections enhance the neutrino opacity and suppress the antineutrino
opacity.  Our calculations, however, indicate that RPA correlations
may significantly suppress the HF neutrino opacity at thermal neutrino
energies in agreement with previous studies~\cite{Reddy_PRC59,Burrows.Sawyer:1999}. For antineutrino
absorption the relative contribution of RPA corrections is energy
dependent: for near threshold antineutrinos they enhance the opacity,
while for higher energies they act in the opposite direction. The
joint effect of mean-field energy shift and long-range RPA
correlations make the protoneutron star matter more opaque
(transparent) for (anti)neutrino radiation as compared to the
noninteracting Fermi gas model.

\begin{acknowledgement}
  This work was supported by the Heisenberg-Landau Program. G.M.-P. is
  partly supported by the Deutsche Forschungsgemeinschaft through
  contract SFB~1245.
\end{acknowledgement}


\end{document}